\documentclass{article}

\usepackage{amsmath,amsfonts}

\usepackage{geometry}
\usepackage{graphicx}

\usepackage{natbib}

\begin{document}

\title{A note on evolutionary stochastic portfolio optimization and probabilistic constraints}
\author{Ronald Hochreiter}




\maketitle

\begin{abstract}
In this note, we extend an evolutionary stochastic portfolio optimization framework to include probabilistic constraints. Both the stochastic programming-based modeling environment as well as the evolutionary optimization environment are ideally suited for an integration of various types of probabilistic constraints. We show an approach on how to integrate these constraints. Numerical results using recent financial data substantiate the applicability of the presented approach.
\end{abstract}


\section{Introduction}
\label{sec:introduction}

Stochastic programming is a powerful method to solve optimization problems under uncertainty, see \cite{RuszczynskiShapiro2003} for theoretical properties and \cite{WallaceZiemba2005} for an overview of possible applications. One specific application area is portfolio optimization, which was pioneered by H.M. Markowitz \cite{Markowitz1952}. The advantage of using the stochastic programming approach is that the optimization can be done without using a covariance matrix of the assets, which is on one hard to estimate and on the other hand does not capture the uncertainty in sufficient detail, especially if there are many assets. 

Instead of using the asset means and the covariance matrix, a stochastic programming approach uses a set of scenarios, each weighted by a certain probability. In the specific portfolio optimization context one scenario is a set of one possible asset return per asset under consideration - see below for more details or e.g. see Chapter 16 of \cite{CornuejolsT2007}. 

This note is organized as follows: Section \ref{sec:espo} summarizes the evolutionary approach, which was used to solve the stochastic portfolio optimization problems. Section \ref{sec:probcons} adds probabilistic constraints to the standard problem and shows an approach on how to integrate these type of constraints 
Section \ref{sec:numres} summarizes numerical results using recent financial data, while Section \ref{sec:conclusion} concludes the note.

\section{Evolutionary Stochastic Portfolio Optimization}
\label{sec:espo}

We follow the approach taken by \cite{Hochreiter2007} and \cite{Hochreiter2008}, which builds a stochastic programming-based environment on top of the general evolutionary portfolio optimization findings reported by \cite{StreichertEtAl03}, \cite{StreichertEtAl04}, and \cite{StreichertEtAl04b}. 

\subsection{Stochastic portfolio optimization}

Let us define the stochastic portfolio optimization problem as follows. We consider a set of assets (or asset categories) $\mathcal{A}$ with cardinality $a$. Furthermore, we base our decision on a scenario set $\mathcal{S}$, which contains a finite number $s$ of scenarios each containing one uncertain return per asset. Each scenario is equipped with a non-negative probability $p_s$, such that $\sum_{s \in \mathcal{S}} p_s = 1$. The scenario set can be composed of historical data or might be the output of a scenario simulator. 

For every portfolio $x$ we can easily calculate the profit and loss distribution by multiplying the scenario matrix with the portfolio weighted by the respective probability. Let us denote the profit function of a certain portfolio $x$ by $\pi(x)$ and the loss function by $\ell(x) = -\pi(x)$.

Every portfolio optimization procedure is a multi-objective optimization. In the traditional case it is a trade-off between return and risk of the profit and loss function. We do not want to employ an multi-objective approach such that we use the classical Markowitz approach and use the Variance of the loss distribution for the risk dimension, which we want to minimize, and the expectation of the profit function for the reward dimension, on which we want to set a lower limit. Hence, the main optimization problem is shown in Eq. (\ref{equ:espo_obj}) below.
\begin{eqnarray}
\begin{array}{ll}
\text{\tt minimize } & \text{Variance}( \ell_x ) \\
\text{\tt subject to } & \mathbb{E}( \pi_x ) > \mu \\
\end{array}
\label{equ:espo_obj}
\end{eqnarray}

Furthermore we consider the classical constraints, i.e. budget normalization, as well as setting an upper and lower limit on each asset position, as shown in Eq. (\ref{equ:espo_cons1}). These are naturally fulfilled by the evolutionary approach shown below, especially since we restrict short-selling in our case.
\begin{eqnarray}
\begin{array}{lll}
\text{\tt subject to } & \sum_{a \in \mathcal{A}} x_a = 1 & \\
& l \leq x_a \leq u & \forall a \in \mathcal{A} \\
\end{array}
\label{equ:espo_cons1}
\end{eqnarray}

\subsection{Evolutionary stochastic portfolio optimization}
\label{sec:espoespo}

The evolutionary algorithm chosen is based on the commonly agreed standard as surveyed by \cite{BlumR03} and is based on the literature cited at the beginning of this Section.

We use the following genetic encoding of a portfolio. Each gene consists of of two parts: One that determines the amount of budget to be distributed to each selected asset and one part which determines in which assets to invest. The first part $g_1$ consists of a predefined number $b$ of real values between $0$ and $1$ and the second part $g_2$ is encoded as a bit-string of the size of the amount of assets. 

The following evolutionary operators have been implemented and used:
\begin{itemize}
\item Elitist selection. A certain number $o_1$ of the best chromosomes will be used within the next population.
\item Crossover. A number $o_2$ of crossovers will be added, both $1$-point crossovers ($g_1$ and $g_2$) and intermediate crossovers (only $g_1$).
\item Mutation. $o_3$ mutated chromosomes will be added. 
\item Random addition. Furthermore $o_4$ randomly sampled chromosomes are added, which are also used for creating the initial population.
\end{itemize}
The specific number of operators $o = (o_1, o_2, o_3, o_4)$ has to be determined for each concrete number of assets $a$ as well as the parameter $b$. 

\section{Probabilistic Constraints}
\label{sec:probcons}

The main advantage of the stochastic programming approach is that the complete distribution is naturally available and can be used for optimization purposes by integrating these directly into the constraint handling mechanism. In the most simplest case, we want to restrict that the probability that the loss exceeds a certain value $\delta$ is lower than a specified probability $\varepsilon$. Given our profit function $\pi_x$, the constraint we want to add to our optimization model is given in Eq. (\ref{equ:espo_cons_prob}).
\begin{equation}
\text{\tt subject to } \mathbb{P}( \pi_x \leq \delta ) \leq \varepsilon.
\label{equ:espo_cons_prob} 
\end{equation}
We will not treat probabilistic constraints as a hard constraint, but use the size of the violation for adding a penalty to the objective function. Let the fitness value which we aim to minimizing be $f$. We calculate the penalty $p$ by
$$p = f \times ( \mathbb{P}( \pi_x \leq \delta ) - \varepsilon ) \times \gamma,$$
where $\gamma$ is a factor to control the penalization level. The fitness value used for evolutionary optimization purposes is thus given by $f' = f + p$. Such a constraint can be implemented and handled conveniently. 

\section{Numerical Results}
\label{sec:numres}

The program code was implemented using MatLab 2008b without using any further toolboxes. 

We use data from the $30$ stocks of the Dow Jones Industrial Average at the beginning of 2010, i.e. the ticker symbols AA, AXP, BA, BAC, CAT, CSCO, CVX, DD, DIS, GE, HD, HPQ, IBM, INTC, JNJ, JPM, KFT, KO, MCD, MMM, MRK, MSFT, PFE, PG, T, TRV, UTX, VZ, WMT, XOM. Daily data from each trading day in 2009 has been used. Weekly returns have been calculated. 

We are using $b = 100$ buckets which are distributed to the respective asset picks, such that each chromosome has a length of $b + a = 130$. The initial population consists of $1000$ random chromosomes. The operator structure defined in Section \ref{sec:espoespo} is $o = (100, 420, 210, 100)$, such that each follow-up population has a size of $830$. This number is due to the different combinations between the crossovers and mutations of $g_1$ and $g_2$.

First, we optimize without using the probabilistic constraints, i.e. the main optimization problem given Eq. \ref{equ:espo_obj} using $\mu = 0.001$. Then we add the probabilistic constraint shown in Eq. \ref{equ:espo_cons_prob} with $\delta = $ and $\varepsilon = 0.1$. These values have been chosen after analysing the resulting loss distribution. 

\begin{figure}
\begin{center}
\begin{tabular}{cc}
\scalebox{0.5}{
\includegraphics{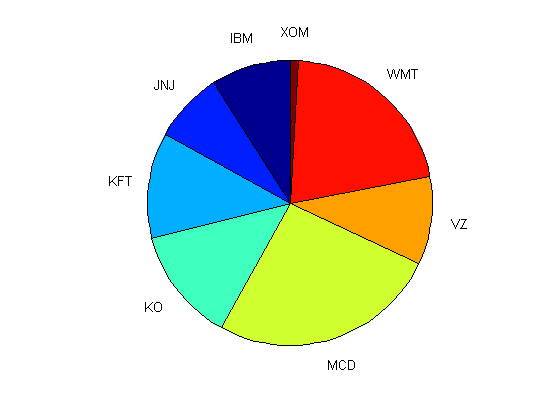}
}
&
\scalebox{0.5}{
\includegraphics{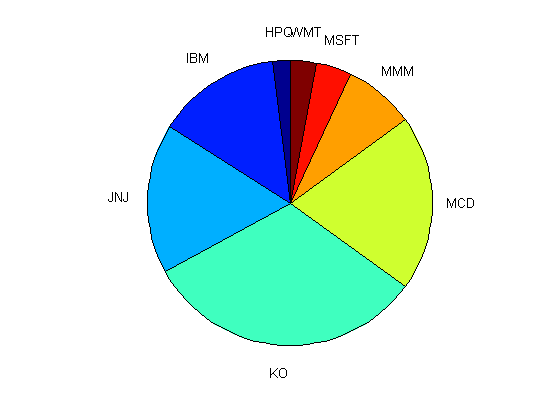}
}
\end{tabular}
\end{center}
\caption{$P_1$ without $\mathbb{P}$ constraint (left) and $P_2$ with $\mathbb{P}$ constraint (right).}
\label{fig:result1}
\end{figure}

\begin{figure}
\begin{center}
\begin{tabular}{cc}
\scalebox{0.5}{
\includegraphics{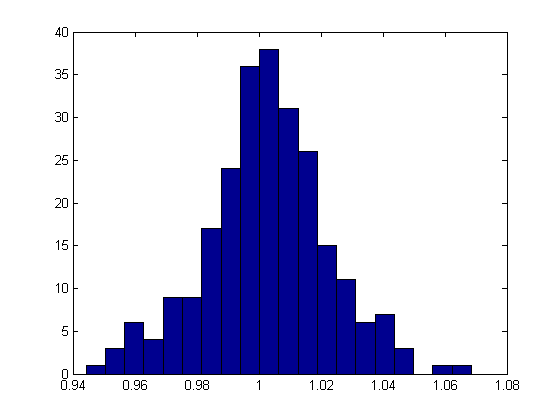}
}
&
\scalebox{0.5}{
\includegraphics{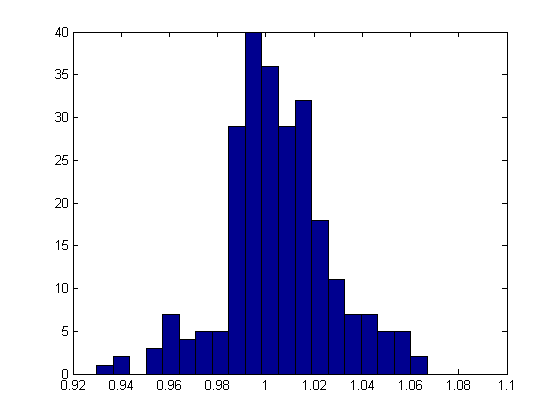}
}
\end{tabular}
\end{center}
\caption{Loss distribution of $P_1$ (left) and $P_2$ (right).}
\label{fig:result2}
\end{figure}

Fig. \ref{fig:result1} shows the optimal portfolio without applying the probabilistic constraint $P_1$ (left) and the optimal one using the probabilistic constraint $P_2$ (right). The allocation changed considerably. The diversification has not changed, i.e. three assets (KFT, VZ, XOM) are dropped from the portfolio, and three others are picked (HPQ, MSFT, MMM). The resulting loss distributions are shown in Fig. \ref{fig:result2}, where the impact of the probabilistic constraint is immediately visible. Furthermore, the statistical properties of the portfolios are shown in Table \ref{tab:prop1}. In this table, the naive $1/N$ portfolio $P_3$ has been added for comparison purposes. While $P_3$ has the highest return, it is also the most risky one in terms of both risk parameters - standard deviation and the probability to fall below the specified threshold. Another interesting fact is that the probabilistic constrained portfolio yields a higher expected return than the standard optimal portfolio. This is partly due to the fact that the lower level $\mu$ has been set to a level below the expected return of the standard portfolio but is definitely another indicator that the plain classical Markowitz approach should not be used for contemporary portfolio optimization purposes. 

\begin{table}
\begin{center}
\begin{tabular}{lccc}
& $P_1$ (no $\mathbb{P}$) & $P_2$ & $P_3 (1/N)$ \\ \hline
Mean & 0.0024 & 0.0051 & 0.0062 \\
Std.Dev. & 0.02 & 0.0225 & 0.0398 \\
Prob. & 0.1774 & 0.1089 & 0.2702 \\
\end{tabular}
\end{center}
\caption{Statistical properties of various portfolios.}
\label{tab:prop1}
\end{table}

\section{Conclusion}
\label{sec:conclusion}

In this note, an extension of an Evolutionary Stochastic Portfolio Optimization to include probabilistic constraints has been presented. It can be shown that the integration of such constraints is straightforward as the underlying probability space is the main object considered for the optimization evaluation. Numerical results visualized the impact of using such constraints in the area of financial engineering. Future extensions of this work include the integration of risk measures into the probabilistic constraint, e.g. constraining the maximum draw-down of the optimal portfolio.

\bibliographystyle{plainnat}
\bibliography{espoprob}

\end{document}